\documentclass[english]{article}
\usepackage[T1]{fontenc}
\usepackage[latin1]{inputenc}
\usepackage{amsmath}
\usepackage{amssymb}

\makeatletter
\newcommand{\lyxaddress}[1]{
\par {\raggedright #1
\vspace{1.4em}
\noindent\par}
}

\usepackage{babel}
\makeatother
\begin{document}

\title{HIGHER DIMENSIONAL SUPERSYMMETRY}

\author{O.P.S. Negi %
\thanks{\noindent Talk presented at conference on, {}``FUNCTION THEORIES
IN HIGHER DIMENSIONS'', Tampere University of Technology, Tampere,
Finland, June12-16, 2006.\protect \\
Address from July 01 to August 31,2006- Universität Konstanz, Fachbereich
Physik, Prof. Dr. H.Dehnen,Postfach M 677, D-78457 Konstanz,Germany%
}}

\maketitle

\lyxaddress{\begin{center}Department of Physics\\
Kumaun University\\
S.~S.~J.~Campus\\
Almora- 263601, U.A., India\\
E-mail:- ops\_negi@yahoo.co.in \par\end{center}}

\begin{abstract}
Higher dimensional super symmetry has been analyzed in terms of quaternion
variables and the theory of quaternion harmonic oscillator has been
analyzed. Supersymmertization of quaternion Dirac equation has been
developed for massless,massive and interacting cases including generalized
electromagnetic fields of dyons. Accordingly higher dimensional super
symmetric gauge theories of dyons are analyzed.
\end{abstract}

\section{Introduction}

Quaternions were very first example of hyper complex numbers having
the significant impacts on mathematics \& physics \cite{key-1} .
Because of their beautiful and unique properties quaternions attached
many to study the laws of nature over the field of these numbers.
Quaternions are already used in the context of special relativity
\cite{key-2}, electrodynamics \cite{key-3,key-4}, Maxwell's equation\cite{key-5},
quantum mechanics\cite{key-6,key-7}, Quaternion Oscillator\cite{key-8},
gauge theorie \cite{key-9,key-10} , Supersymmetry \cite{key-11}and
many other branches of Physics\cite{key-12} and Mathematics\cite{key-13}.On
the other hand supersymmetry (SUSY) is described as the symmetry of
bosons and fermions\cite{key-14,key-15,key-16}. Gauge Hierarchy problem,
not only suggests that the SUSY exists but put an upper limit on the
masses of super partners\cite{key-17,key-18}. The exact SUSY implies
exact fermion-boson masses, which has not been observed so far. Hence
it is believed that supersymmetry is an approximate symmetry and it
must be broken \cite{key-19,key-20}. We have considered following
two motivations to study the higher dimensional supersymmetric quantum
mechanics\cite{key-21} over the field of Quaternions. 

1. Supersymmetric field theory can provide us realistic models of
particle physics which do not suffer from gauge hierarchy problem
and role of quaternions will provide us simplex and compact calculation
accordingly.

2. Quaternions super symmetric quantum mechanics can give us new window
to understand the behavior of supersymmetric partners and mechanism
of super symmetry breaking etc.

3. Quaternions are capable to deal the higher dimensional structure
and thus include the theory of monopoles and dyons.

Keeping these facts in mind and to observe the role of quaternions
in supersymmetry, the theory of quaternion harmonic oscillator has
been analyzed for the systems of bosons and fermions respectively
in terms of commutation and anti commutation relations. Eigen values
of particle Hamiltonian and number operators are calculated by imposing
the restriction on the component of quaternion variables. Accordingly,
the super charges are calculated and it is shown that the Hamiltonian
operator commutes with the super charges representing the conversion
of a fermionic state to a bosonic state and vice versa. Quaternion
reformulation of N= 1 , 2 and 4 dimension supersymmetry has been investigated
in terms of supercharges and super partner potential and quaternion
mechanics has also been analyzed in terms of complex and quaternion
quantum mechanics for N = 2 and N = 4 SUSY respectively. It has been
shown that elegant frame work of quaternion quantum mechanics includes
non Abelian gauge structure in contradiction to complex quantum mechanics
of supersymmetry corresponding to N= 2, complex and N= 4 real dimension
of supersymmetry.

\section{Definition}

A quaternion $\phi$ is expressed as

\begin{eqnarray}
\phi & = & e_{\text{0}}\phi_{0}+e_{1}\phi_{1}+e_{2}\phi_{2}+e_{3}\phi_{3}\label{eq:1}\end{eqnarray}

Where $\phi_{0},\phi_{1},\phi_{2},\phi_{3}$are the real quartets
of a quaternion and $e_{0},e_{1},e_{2},e_{3}$ are called quaternion
units and satisfies the following relations,

\begin{eqnarray}
e_{0}^{2} & = & e_{0}=1\nonumber \\
e_{0}e_{i} & = & e_{i}e_{0}=e_{i}(i=1,2,3)\nonumber \\
e_{i}e_{j} & = & -\delta_{ij}+\varepsilon_{ijk}e_{k}(i,j,k=1,2,3)\label{eq:2}\end{eqnarray}

The quaternion conjugate $\bar{\phi}$ is then defined as \begin{eqnarray}
\bar{\phi} & = & e_{\text{0}}\phi_{0}-e_{1}\phi_{1}-e_{2}\phi_{2}-e_{3}\phi_{3}\label{eq:3}\end{eqnarray}

Here $\phi_{0}$is real part of the quaternion defined as\begin{align}
\phi_{0} & =Re\,\,\phi=\frac{1}{2}(\bar{\phi}+\phi)\label{eq:4}\end{align}
 $Re\,\,\phi=\phi_{0}=0$ , then $\phi=-\bar{\phi}$ and imaginary
$\phi$ is called pure quaternion and is written as

\begin{eqnarray}
Im\,\,\phi & = & e_{1}\phi_{1}+e_{2}\phi_{2}+e_{3}\phi_{3}\label{eq:5}\end{eqnarray}

The norm of a quaternion is expressed as 

\begin{eqnarray}
N(\phi) & = & \bar{\phi}\phi=\phi\bar{\phi}=\phi_{0}^{2}+\phi_{1}^{2}+\phi_{2}^{2}+\phi_{3}^{2}\geq0\label{eq:6}\end{eqnarray}

and the inverse of a quaternion is described as

\begin{eqnarray}
\phi^{-1} & = & \frac{\bar{\phi}}{\left|\phi\right|}\label{eq:7}\end{eqnarray}

While the quaternion conjugation satisfies the following property

\begin{eqnarray}
\overline{(\phi_{1}\phi_{2})} & = & \bar{\phi_{2}}\bar{\phi}_{1}\label{eq:8}\end{eqnarray}

The norm of the quaternion (\ref{eq:6}) is positive definite and
enjoys the composition law

\begin{eqnarray}
N(\phi_{1}\phi_{2}) & = & N(\phi_{1})N(\phi_{2})\label{eq:9}\end{eqnarray}

Quaternion (\ref{eq:1}) is also written as $\phi=(\phi_{0},\overrightarrow{\phi})$where
$\overrightarrow{\phi}=$$e_{1}$$\phi_{1}+e_{2}\phi_{2}+e_{3}$$\phi_{3}$
is its vector part and $\phi_{0}$is its scalar part. The sum and
product of two quaternions are

\begin{eqnarray}
(\alpha_{0},\overrightarrow{\alpha})+(\beta_{0},\overrightarrow{\beta}) & = & (\alpha_{0}+\beta_{0},\overrightarrow{\alpha}+\overrightarrow{\beta})\nonumber \\
(\alpha_{0},\overrightarrow{\alpha})\,(\beta_{0},\overrightarrow{\beta}) & = & (\alpha_{0}\beta_{0}-\overrightarrow{\alpha}.\overrightarrow{\beta},\,\alpha_{0}\overrightarrow{\beta}+\beta_{0}\overrightarrow{\alpha}+\overrightarrow{\alpha}\times\overrightarrow{\beta})\label{eq:10}\end{eqnarray}

Quaternion elements are non-Abelian in nature and thus represent a
division ring.

\section{Field Associated with Dyons}

Let us define the generalized charge on dyons as \cite{key-22,key-23},

\begin{eqnarray}
q & = & q=e\,\,-\,\, i\,\, g\,\,\,(i=\sqrt{-1})\label{eq:11}\end{eqnarray}

where e and g are respectively electric and magnetic charges. Generalized
four potential $\left\{ V_{\mu}\right\} =\left(\varphi,\overrightarrow{V}\right)$associated
with dyons is defined as,

\begin{eqnarray}
\left\{ V_{\mu}\right\}  & = & \left\{ A_{\mu}\right\} -\, i\,\,\left\{ B_{\mu}\right\} =(\varphi,\overrightarrow{V)}\label{eq:12}\end{eqnarray}

where $\left\{ A_{\mu}\right\} =\left(\varphi_{e},\overrightarrow{A}\right)$and
$\left\{ B_{\mu}\right\} =\left(\varphi_{g},\overrightarrow{B}\right)$are
respectively electric and magnetic four potentials. We have used throughout
the natural units $c=\hslash=1$. Electric and magnetic fields of
dyons are defined in terms of components of electric and magnetic
potentials as,

\begin{eqnarray}
\overrightarrow{E} & = & -\frac{\partial\overrightarrow{A}}{\partial t}-\overrightarrow{\nabla}\varphi_{e}-\overrightarrow{\nabla}\times\overrightarrow{A}\nonumber \\
\overrightarrow{H} & = & -\frac{\partial\overrightarrow{B}}{\partial t}-\overrightarrow{\nabla}\varphi_{g}-\overrightarrow{\nabla}\times\overrightarrow{B}\label{eq:13}\end{eqnarray}

These electric and magnetic fields of dyons are invariant under generalized
duality transformation i.e. 

\begin{eqnarray}
\left\{ A_{\mu}\right\}  & \Rightarrow & \left\{ A_{\mu}\right\} cos\theta+\left\{ B_{\mu}\right\} sin\theta\nonumber \\
\left\{ B_{\mu}\right\}  & \Rightarrow & -\left\{ A_{\mu}\right\} sin\theta+\left\{ B_{\mu}\right\} cos\theta\label{eq:14}\end{eqnarray}

The expression of generalized electric and magnetic fields given by
equation (\ref{eq:13}) are symmetrical and both the electric and
magnetic fields of dyons may be written in terms of longitudinal and
transverse components. The generalized vector electromagnetic fields
associated with dyons is defined as 

\begin{eqnarray}
\overrightarrow{\psi} & = & \overrightarrow{E}-i\,\,\overrightarrow{H}\label{eq:15}\end{eqnarray}

As such, we get the following differential form of generalized Maxwell's
equations for dyons i.e.

\begin{eqnarray}
\overrightarrow{\nabla}\cdot\overrightarrow{\psi} & = & J_{0}\nonumber \\
\overrightarrow{\nabla}\times\overrightarrow{\psi} & = & -i\,\,\overrightarrow{J}-i\,\,\frac{\partial\overrightarrow{\psi}}{\partial t}\label{eq:16}\end{eqnarray}
where $j_{0}$ and $\overrightarrow{J}$, are the generalized charge
and current source densities of dyons, given by;

\begin{eqnarray}
\left\{ J_{\mu}\right\}  & = & \left\{ j_{\mu}\right\} -\,\, i\,\,\left\{ k_{\mu}\right\} =\left(J_{0},\overrightarrow{J}\right)\label{eq:17}\end{eqnarray}

Substituting relation (\ref{eq:13}) into equation (\ref{eq:15})
and using equation (\ref{eq:12}), we obtain the following relation
between generalized vector field and potential of dyons i.e.

\begin{eqnarray}
\overrightarrow{\psi} & = & -\frac{\partial\overrightarrow{V}}{\partial t}-\overrightarrow{\nabla}\varphi-i\,\,\overrightarrow{\nabla}\times\overrightarrow{V}\label{eq:18}\end{eqnarray}

where electric and magnetic four current densities$\left\{ j_{\mu}\right\} =(\rho_{e},\,\overrightarrow{j}$)and
$\left\{ k_{\mu}\right\} =(\rho_{g},$$\overrightarrow{k}\,)$Thus
we write the following tensor forms of generalized Maxwell's-Dirac
equations of dyons i.e.

\begin{eqnarray}
F_{\mu\nu,\,\nu} & = & \partial^{\nu}F_{\mu\nu}=j_{\mu}\nonumber \\
\tilde{F}_{\mu\nu,\,\nu} & = & \partial^{\nu}\tilde{F}_{\mu\nu}=k_{\mu}\label{eq:19}\end{eqnarray}
where

\begin{eqnarray}
F_{\mu\nu} & = & E_{\mu\nu}-\tilde{H}_{\mu\nu}\nonumber \\
\tilde{F}_{\mu\nu} & = & H_{\mu\nu}+\tilde{E}_{\mu\nu}\label{eq:20}\end{eqnarray}

and

\begin{eqnarray}
E_{\mu\nu} & = & A_{\mu,\nu}-A_{\nu,\mu}\nonumber \\
H_{\mu\nu} & = & B_{\mu,\nu}-B_{\nu,\mu}\nonumber \\
\tilde{E}_{\mu\nu} & = & \frac{1}{2}\varepsilon_{\mu\nu\sigma\lambda}E^{\sigma\lambda}\nonumber \\
\tilde{H}_{\mu\nu} & = & \frac{1}{2}\varepsilon_{\mu\nu\sigma\lambda}H^{\sigma\lambda}\label{eq:21}\end{eqnarray}

The tidle denotes the dual part while $\varepsilon_{\mu\nu\sigma\lambda}$
are four indexes Levi-Civita symbol. Generalized fields of dyons given
by equation (\ref{eq:13}) may directly be obtained from field tensors
$F_{\mu\nu}$and $\tilde{F}_{\mu\nu}$ as,

\begin{eqnarray}
F_{0i} & = & E^{\, i\,}\nonumber \\
F_{ij} & = & \varepsilon_{ijk}H^{\, k}\nonumber \\
\tilde{H}_{0i} & = & -H^{\, i}\nonumber \\
\tilde{H}_{ij} & = & -\varepsilon_{ijk}E^{^{k}}\label{eq:22}\end{eqnarray}

A new vector parameter S (say) may directly be obtained \cite{key-10,key-12}from
equation (\ref{eq:16})i.e.

\begin{eqnarray}
\overrightarrow{S} & = & \square\overrightarrow{\psi}=-\frac{\partial\overrightarrow{\psi}}{\partial t}-\overrightarrow{\nabla}\rho-i\,\,\overrightarrow{\nabla}\times\overrightarrow{\psi}\label{eq:23}\end{eqnarray}

where $\square$represents the D' Alembertian operator i.e.

\begin{eqnarray}
\square & = & \frac{\partial^{2}}{\partial t\,^{2}}-\frac{\partial^{2}}{\partial x\,^{2}}-\frac{\partial^{2}}{\partial y^{\,2}}-\frac{\partial^{2}}{\partial z\,^{2}}\label{eq:24}\end{eqnarray}

Defining generalized field tensor as 

\begin{eqnarray}
G_{\mu\nu} & = & F_{\mu\nu}-i\,\tilde{F}_{\mu\nu}\label{eq:25}\end{eqnarray}

We may directly obtain the following generalized field equation of
dyons i.e.

\begin{eqnarray}
G_{\mu\nu,\,\nu} & = & \partial^{\nu}G_{\mu\nu}=J_{\mu}\label{eq:26}\end{eqnarray}

where $G_{\mu\nu}=V_{\mu,\nu}-V_{\nu,\mu}$is called the generalized
electromagnetic field tensor of dyons. Equation (\ref{eq:26}) may
also be written as follows like second order Klein-Gordon equation
for dyonic fields;

\begin{eqnarray}
\square V_{\mu} & = & J_{\mu}\label{eq:27}\end{eqnarray}

where we have imposed the Lorentz gauge condition on both potentials
and consequently to generalized potential. Equations ( \ref{eq:19})
and (\ref{eq:26}) are also invariant under duality transformations;

\begin{eqnarray}
(F,\tilde{F}) & = & (F\, cos\,\theta+\tilde{F}sin\,\theta;-F\, sin\,\theta+\,\tilde{F}\, cos\,\theta)\label{eq:28}\\
(j_{\mu},k_{\mu}) & = & (j_{\mu}cos\,\theta+k_{\mu}sin\,\theta;-j_{\mu}sin\,\theta+k_{\mu}cos\,\theta)\label{eq:29}\end{eqnarray}

where 

\begin{eqnarray}
\frac{g}{e} & = & \frac{B_{\mu}}{A_{\mu}}=\frac{k_{\mu}}{j_{\mu}}=-tan\,\theta=Constant\label{eq:30}\end{eqnarray}
and consequently the generalized charge of a dyon may be written as 

\begin{eqnarray}
q & = & \left|q\right|exp\,(-i\,\theta)\label{eq:31}\end{eqnarray}

The suitable Lagrangian density, which yields the field equation (\ref{eq:26})
under the variation of field parameters (i.e. potential only) without
changing the trajectory of particle, may be written as follows;

\begin{eqnarray}
\mathit{\mathsf{\mathbf{L}}} & = & -\frac{1}{4}G_{\mu\nu}^{*}G_{\mu\nu}+V_{\mu}^{*}V_{\mu}\label{eq:32}\end{eqnarray}
where {*} denotes the complex conjugate. Lagrangian density given
by equation (\ref{eq:32}) directly leads the following expression
of Lorentz force equation of motion for dyons i.e.

\begin{eqnarray}
f_{\mu} & = & Re\,(q^{*}G_{\mu\nu})u^{\nu}\label{eq:33}\end{eqnarray}
Where Re denotes real part and $\left\{ u^{\nu}\right\} $ is the
four-velocity of the particle.

\section{Quaternion SUSY Harmonic Oscillator}

Let us define bosonic quaternion oscillator as the extension of complex
oscillator having the decomposition \cite{key-8}as,

\begin{eqnarray}
\hat{a} & = & \frac{1}{\sqrt{6}}\left[\widehat{a}_{0}+e_{1}\widehat{a}_{1}+e_{2}\widehat{a}_{2}+e_{3}\widehat{a}_{3}\right]\label{eq:34}\end{eqnarray}
where $\widehat{a}_{0},\widehat{a}_{1},\widehat{a}_{2},\widehat{a}_{3}$are
real operators . Let us defined the conjugate of equation (\ref{eq:34})
as

\begin{eqnarray}
\hat{a}^{\dagger} & = & \frac{1}{\sqrt{6}}\left[\widehat{a}_{0}-e_{1}\widehat{a}_{1}-e_{2}\widehat{a}_{2}-e_{3}\widehat{a}_{3}\right]\label{eq:35}\end{eqnarray}
Like other oscillator, we start with the following fundamental boson
commutation relations i.e.

\begin{eqnarray}
\left[\hat{a}\,\,,\,\,\hat{a}^{\dagger}\right] & = & 1\nonumber \\
\left[\hat{a}\,\,,\,\,\hat{a}\right] & = & 0\nonumber \\
\left[\hat{a}^{\dagger},\,\hat{a}^{\dagger}\right] & = & 0\label{eq:36}\end{eqnarray}
Then to maintain the above relations of bosonic oscillator,we get
the following commutation relation between the components of bosonic
oscillator in terms of imaginary quaternion units i.e.

\begin{eqnarray}
\left[\hat{a}_{0}\,\,,\,\,\hat{a_{1}}\right] & = & e_{1}\nonumber \\
\left[\hat{a}_{0}\,\,,\,\,\hat{a_{2}}\right] & = & e_{2}\nonumber \\
\left[\hat{a}_{0}\,\,,\,\,\hat{a}_{3}\right] & = & e_{3}\label{eq:37}\end{eqnarray}

or\begin{eqnarray}
\left[\hat{a}_{0}\,\,,\,\,\hat{a_{A}}\right] & = & e_{A}(A=1,2,3)\nonumber \\
\left[\hat{a}_{\mu}\,\,,\,\,\hat{a}_{\nu}\right] & = & \,\,0\quad\forall\,\,\,\mu\neq\nu\label{eq:38}\end{eqnarray}

Let us describe the Hamiltonian for bosonic harmonic oscillator as

\begin{eqnarray}
\hat{H_{B}} & = & \frac{p^{2}}{2m}+\frac{1}{2}m\omega^{2}q^{2}\label{eq:39}\end{eqnarray}

which can be written in terms of as 

\begin{eqnarray}
\hat{H_{B}} & = & \frac{1}{2}\hbar\omega(\hat{a}\,\hat{a}^{\dagger}+\hat{a}^{\dagger}\hat{a)}=\hbar\omega(\hat{a}^{\dagger}\hat{a\,+\frac{1}{2})}\label{eq:40}\end{eqnarray}
 where 

\begin{eqnarray}
\hat{a} & = & \frac{1}{\sqrt{2m\hbar\omega}}\left(m\omega\hat{q}-e_{1}\hat{p}\right)\nonumber \\
\hat{a}^{\dagger} & = & \frac{1}{\sqrt{2m\hbar\omega}}\left(m\omega\hat{q}+e_{1}\hat{p}\right)\label{eq:41}\end{eqnarray}

Then it is necessary to recover the ordinary commutation relation
$\left[\hat{q},\hat{p}\right]=i$ $\hbar$ between $\hat{q}\hat{,p}$
.So, we get

\begin{eqnarray}
\hat{p} & = & \sqrt{\frac{m\hbar\omega}{3}}\left(-\widehat{a_{1}}+e_{3}\widehat{a_{2}}-e_{2}\widehat{a_{3}}\right)\nonumber \\
\widehat{q} & = & \widehat{a_{0}}\sqrt{\frac{\hbar}{3m\omega}}\label{eq:42}\end{eqnarray}

Now we can define the number operator as 

\begin{eqnarray}
\widehat{N_{B}} & = & \widehat{a}\,^{\dagger}\widehat{a}\label{eq:43}\end{eqnarray}

which thus commutes with Hamiltonian $\hat{H_{B}}$i.e.

\begin{eqnarray}
\left[\widehat{N_{B}}\,\,,\,\,\widehat{H_{B}}\right] & = & 0\label{eq:44}\end{eqnarray}

Bosonic number operator satisfies the following relations;

\begin{eqnarray}
\left[\widehat{N_{B}}\,\,,\,\,\widehat{a}\,^{\dagger}\right] & = & \widehat{a}^{\dagger}\nonumber \\
\left[\widehat{N_{B}}\,\,,\,\,\widehat{a}\,\right] & = & -\widehat{a}\label{eq:45}\end{eqnarray}
 which shows that $\hat{a}$ and $\hat{a}^{\dagger}$may be regarded
as annihilation and creation operators. We also have 

\begin{eqnarray}
\hat{a}{\displaystyle \left|0\right\rangle } & = & 0\nonumber \\
\hat{a}^{\dagger}\left|n\right\rangle  & = & \sqrt{n+1}\left|n+1\right\rangle \nonumber \\
\hat{a}{\displaystyle \left|n\right\rangle } & = & \sqrt{n}\left|n-1\right\rangle \nonumber \\
\widehat{N_{B}}\left|n\right\rangle  & = & n\left|n\right\rangle \label{eq:46}\end{eqnarray}

and the Hilbert space is then spanned by state vectors as 

\begin{eqnarray}
\left|n\right\rangle  & = & \frac{1}{\sqrt{n}!}(\hat{a}^{\dagger})^{n}\left|n\right\rangle \label{eq:47}\end{eqnarray}

where $\left|0\right\rangle $ is considered as ground or vacuum state
and then must be normalized as

\begin{eqnarray}
\left\langle 0\mid0\right\rangle  & = & 1\label{eq:48}\end{eqnarray}

and thus gives rise to the familiar results

\begin{eqnarray}
\widehat{H_{B}}\,\left|n\right\rangle  & = & E_{n}\left|n\right\rangle =(n+\frac{1}{2})\left|n\right\rangle ;\,\,\,\, E_{n}=(n+\frac{1}{2})\label{eq:49}\end{eqnarray}

Similarly we can write the following anti commutation relation for
fermionic harmonic oscillator

\begin{eqnarray}
\left\{ \widehat{b\,},\widehat{b}^{\dagger}\right\}  & = & 1\nonumber \\
\left\{ \widehat{b\,},\widehat{b}\right\}  & = & 0\nonumber \\
\left\{ \widehat{b\,}^{\dagger},\widehat{b}^{\dagger}\right\}  & = & 0\label{eq:50}\end{eqnarray}

where $\widehat{b}$ is a fermionic quaternion operator and may be
decomposed as

\begin{eqnarray}
\hat{b} & = & \frac{1}{\sqrt{6}}\left[\widehat{b}_{0}+e_{1}\widehat{b}_{1}+e_{2}\widehat{b}_{2}+e_{3}\widehat{b}_{3}\right]\nonumber \\
\hat{b}^{\dagger} & = & \frac{1}{\sqrt{6}}\left[\widehat{b}_{0}-e_{1}\widehat{b}_{1}-e_{2}\widehat{b}_{2}-e_{3}\widehat{b}_{3}\right]\label{eq:51}\end{eqnarray}

To satisfy the relations (\ref{eq:50}) by fermion operators given
by equation (\ref{eq:51}) , it is necessary to impose the following
restrictions on the various components of operators i.e.,

\begin{eqnarray}
b_{0}^{2}+b_{1}^{2} & +b_{2}^{2}+b_{3}^{2}= & 0\nonumber \\
b_{0}^{2}- & b_{1}^{2}-b_{2}^{2}-b_{3}^{2}= & 3\nonumber \\
\left[\widehat{b_{j}}\,,\,\widehat{b}_{k}\right] & =\varepsilon_{jkl} & \widehat{b}_{l}\nonumber \\
\left[\widehat{b}\,,\,\widehat{b}_{j}\right] & =0 & \forall\,\, j,k,l=1,2,3\label{eq:52}\end{eqnarray}
We have defined through out the text $\left[\,,\,\right]$ as commutator,
$\left\{ \,,\,\right\} $as anti commutator and $\varepsilon_{jkl}$
is three index Levi- Civita symbol. Like bosonic harmonic oscillator
let us define the fermionic Hamiltonian as ,

\begin{eqnarray}
\hat{H_{F}} & = & \frac{1}{2}\hbar\omega(\hat{b}\,\hat{b}^{\dagger}+\hat{b}^{\dagger}\hat{b)}=\hbar\omega(\,\hat{b}\,\hat{b}^{\dagger}-\frac{1}{2})=\hbar\omega(\widehat{N_{F}}\,\,-\frac{1}{2})\label{eq:53}\end{eqnarray}

Where $\widehat{N_{F}}\,=\hat{b}^{\dagger}\hat{b}$ and $\widehat{N_{F}}\,$can
take eigen values $n_{f}=0,1$.The Hilbert space with basis vector
$\left|n\right\rangle $ is now constructed in the following manner
so that

\begin{eqnarray}
\widehat{N_{F}}\, & \left|n\right\rangle  & =n_{f}\left|n\right\rangle \,\,\,\,\,\,(n_{f}=0,1)\nonumber \\
\widehat{b} & \left|1\right\rangle  & =\left|0\right\rangle \nonumber \\
\widehat{b}^{\dagger} & \left|0\right\rangle  & =\left|1\right\rangle \nonumber \\
\widehat{b} & \left|0\right\rangle  & =0=\widehat{b}^{\dagger}\left|1\right\rangle \label{eq:54}\end{eqnarray}

The energy eigen spectra of fermionic oscillator have only two levels
for eigen state $\left|0\right\rangle $or $\left|1\right\rangle $
i.e. $E_{0}=-\frac{1}{2}\hbar\omega$ and $E_{1}=+\frac{1}{2}\hbar\omega$showing
that ground state energy of this oscillator is negative i.e. $E_{0}=-\frac{1}{2}\hbar\omega$
.

Let us now construct a simple supersymmetric quantum mechanical system
that include an oscillator with the bosonic and fermionic degrees
of freedom .We call it as supersymmetric harmonic oscillator viewed
in the frame work of quaternion variables. The supersymmetry is thus
obtained by annihilating simultaneously one bosonic quantum and creating
one fermionic quantum or vice versa. We illustrate the annihilating
(supersymmetric) charges (generators) as\begin{eqnarray}
\widehat{Q}= & \sqrt{\hbar\omega} & (\widehat{a}^{\dagger}\widehat{b})\nonumber \\
\widehat{Q}^{\dagger}= & \sqrt{\hbar\omega} & (\widehat{b}^{\dagger}\widehat{a})\label{eq:55}\end{eqnarray}

So the SUSY Hamiltonian becomes

\begin{eqnarray}
\widehat{H} & =\left\{ \widehat{Q}^{\dagger}\,,\,\widehat{Q}\right\}  & =\hbar\omega\left\{ \widehat{a}^{\dagger}\,\widehat{a}\,+\widehat{b}^{\dagger}\widehat{b}\right\} =\hbar\omega(\widehat{N}_{B}+\widehat{N}_{F})\label{eq:56}\end{eqnarray}

and

\begin{eqnarray}
\left[\widehat{H}\,\,,\,\,\widehat{Q}\right] & = & 0=\left[\widehat{H}\,\,,\,\,\widehat{Q}^{\dagger}\right]\label{eq:57}\end{eqnarray}

where $\mathbf{\widehat{N}_{B}}$ and $\widehat{N}_{F}$ are respectively
bosonic and fermionic number operators. The eigen state is described
as$\left|n_{B}\,,\, n_{F}\right\rangle $ and ground state as $\left|0\,.\,0\right\rangle $so
that

\begin{eqnarray}
\widehat{H} & \left|n_{B}\,,\, n_{F}\right\rangle  & =E_{n_{B}\,,\, n_{B}}\left|n_{B}\,,\, n_{F}\right\rangle ;\,\, n_{B}=0,1,2,3,........;n_{F}=0,1.\label{eq:58}\end{eqnarray}

and we also have 

\begin{eqnarray}
\widehat{Q}\left|n,\,1_{F}\right\rangle  & = & \sqrt{n+1}\left|n+1\,,\,0\right\rangle \nonumber \\
\widehat{Q}^{\dagger}\left|n+1,\,0_{F}\right\rangle  & = & \sqrt{n+1}\left|n\,,\,1\right\rangle \label{eq:59}\end{eqnarray}

These supercharges represent conversion of a fermionic state to a
bosonic state and bosonic state to fermionic state or vice versa i.e.

\begin{eqnarray}
\widehat{Q}^{\dagger}\left|BOSON\right\rangle  & = & \left|FERMION\right\rangle \nonumber \\
\widehat{Q}\left|FERMION\right\rangle  & = & \left|BOSON\right\rangle \label{eq:60}\end{eqnarray}

Equations (\ref{eq:56}and \ref{eq:57}) are analogous to following
equations of supersymmetry,

\begin{eqnarray}
\left\{ \widehat{Q_{\alpha}}^{\dagger},\widehat{Q}_{\beta}\right\}  & = & P^{\mu}(\sigma_{\mu})_{\alpha\beta}\label{eq:61}\\
\left[\widehat{H}\,\,,\widehat{Q}_{\alpha}\right] & = & 0\label{eq:62}\end{eqnarray}

for $\mu=0$ and $\alpha=\beta=1$namely one dimensional SUSY .Supercharges
always commute with the usual Hamiltonian.Thus the anti commuting
charges in quaternion formalism combine to form the generators of
time translation, namely the Hamiltonian $H$. The ground state of
this system is the state $\left|0\right\rangle _{osc}.\left|0\right\rangle _{spin}$
or $\left|0\right\rangle _{boson}.\left|0\right\rangle _{fermion}=\left|0\,\,,\,\,0\right\rangle $where
both bosonic and fermionic degrees of freedom are in the lowest energy
state.As such we have analyzed the theory of supersymmetric harmonic
oscillator for one dimensional supersymmetric quantum mechanics and
putting the restriction accordingly. Other wise one has to extend
the dimensions and to loose the hermiticity of the Hamiltonian of
the supersymmetric system.Let us now try to illustrate the Supersymmetry
of Dirac equation in terms of quaternion variables.

\section{Supersymmetric Quaternion Dirac Equation }

The quaternion formulation of free particle Dirac equation\cite{key-6,key-24,key-25}
is described as,

\begin{eqnarray}
i\,\,\gamma_{\mu\,\,}\partial_{\mu}\psi(x,t) & = & m\,\psi(x,t)\,\,(\mu=0,1,2,3)\label{eq:63}\end{eqnarray}

Let us discuss the supersymmetrization in terms of following three
cases.

Case I: For mass less free particle i.e.$m=0$ and external potential
$\Phi=0$;

Equation (\ref{eq:63} )becomes 

\begin{eqnarray}
i\,\,\gamma_{\mu\,\,}\partial_{\mu}\psi(x,t) & = & 0\label{eq:64}\end{eqnarray}

Let us consider the following solutions of this equation as

\begin{eqnarray}
\psi(x,t) & = & \psi(x)\, e^{i\,(\overrightarrow{p}\cdot\overrightarrow{x}-Et)}\label{eq:65}\end{eqnarray}

where we have taken natural units $c=\hbar=1$ and as such we get
the following form of equation

(\ref{eq:64})i.e.\begin{eqnarray}
(\gamma_{0}E\,-\gamma_{1}p_{1}-\gamma_{2}p_{2}-\gamma_{3}p_{3}) & \psi(x)= & 0\label{eq:66}\end{eqnarray}

We define the following representation of gamma matrices in terms
of quaternion units i.e.

\begin{eqnarray}
\gamma_{0} & = & \left[\begin{array}{cc}
1 & 0\\
0 & -1\end{array}\right]\nonumber \\
\gamma_{l} & = & e_{l}\,\left[\begin{array}{cc}
0 & 1\\
1 & 0\end{array}\right]\,\,\,\,(l=1,2,3)\label{eq:67}\end{eqnarray}

thus equation(\ref{eq:66}) takes the form

\begin{eqnarray}
\left\{ \left[\begin{array}{cc}
1 & 0\\
0 & -1\end{array}\right]E-e_{l}\left[\begin{array}{cc}
0 & 1\\
1 & 0\end{array}\right]p_{l}\right\} \left\{ \begin{array}{c}
\psi_{a}\\
\psi_{b}\end{array}\right\}  & = & 0\label{eq:68}\end{eqnarray}

where $\psi_{a}=\psi_{0}+i\,\psi_{1}$and $\psi_{b}=\psi_{2}-i\,\psi_{3}$.We
thus obtain the following coupled equations

\begin{eqnarray}
\widehat{A}\psi_{a}(x) & = & E\,\psi_{b}(x)\nonumber \\
\widehat{A}^{\dagger}\psi_{b}(x) & = & E\,\psi_{a}(x)\label{eq:69}\end{eqnarray}

where $\widehat{A}\,=-e_{l}\,\widehat{p}_{l}$ and $\widehat{A}^{\dagger}=e_{l}\,\widehat{p}_{l}$.We
can now decouple equation (\ref{eq:69}) as 

\begin{eqnarray}
\widehat{A}\,\widehat{A}^{\dagger}\psi_{b}(x) & = & E^{2}\psi_{b}(x)\nonumber \\
\widehat{A}^{\dagger}\widehat{A}\,\psi_{a}(x) & = & E^{2}\psi_{a}(x)\nonumber \\
P_{l}^{2}\psi_{b}(x) & = & E^{2}\psi_{b}(x)\nonumber \\
P_{l}^{2}\psi_{a}(x) & = & E^{2}\psi_{a}(x)\label{eq:70}\end{eqnarray}

where $\psi_{a}(x)$ and $\psi_{b}(x)$are eigen functions of partner
Hamiltonians $H_{-}=\widehat{A}^{\dagger}\widehat{A}$ and $H_{+}=\widehat{A}\,\widehat{A}^{\dagger}$
.The supersymmetric Hamiltonian is thus described as

\begin{eqnarray}
\widehat{H} & = & \left[\begin{array}{cc}
H_{+} & 0\\
0 & H_{-}\end{array}\right]=\left[\begin{array}{cc}
-P_{^{l}}^{2} & 0\\
0 & P_{^{l}}^{2}\end{array}\right]\label{eq:71}\end{eqnarray}

Restricting the propagation along x-axis to discuss the quantum mechanics
in two dimensional space time,we have

\begin{eqnarray}
\widehat{p}_{l} & = & -i\,\frac{d}{dx}\nonumber \\
e_{l}\,\widehat{p}_{l} & = & \frac{d}{dx}\nonumber \\
\widehat{H} & = & \left[\begin{array}{cc}
-\frac{d^{2}}{dx^{2}} & 0\\
0 & -\frac{d^{2}}{dx^{2}}\end{array}\right]\label{eq:72}\end{eqnarray}

or\begin{eqnarray}
\widehat{H} & = & \left[\begin{array}{cc}
\widehat{Q}\widehat{Q}^{\dagger} & 0\\
0 & \widehat{Q}^{\dagger}\widehat{Q}\end{array}\right]\label{eq:73}\end{eqnarray}

where supercharges are described in terms of quaternion units i.e.

\begin{eqnarray}
\widehat{Q} & = & \left[\begin{array}{cc}
0 & -e_{2}^{\dagger}\frac{d}{dx}\\
0 & 0\end{array}\right]\nonumber \\
\widehat{Q}^{\dagger} & = & \left[\begin{array}{cc}
0 & 0\\
-e_{2}^{\dagger}\frac{d}{dx} & 0\end{array}\right]\label{eq:74}\end{eqnarray}

As such we may obtain the supersymmetry algebra as 

\begin{eqnarray}
\left[\widehat{Q\,},\widehat{H}\right] & = & \left[\widehat{Q\,},\widehat{H}^{\dagger}\right]=0\nonumber \\
\left\{ \widehat{Q\,},\widehat{Q\,}\right\}  & = & \left\{ \widehat{Q\,}^{\dagger},\widehat{Q\,}^{\dagger}\right\} =0\nonumber \\
\left\{ \widehat{Q\,},\widehat{Q\,}^{\dagger}\right\}  & = & \widehat{H}\label{eq:75}\end{eqnarray}

Here $\widehat{Q},$converts the upper component spinor $\left\{ \begin{array}{c}
\psi_{a}\\
0\end{array}\right\} $to a lower one $\left\{ \begin{array}{c}
0\\
\psi_{b}\end{array}\right\} $and convert the lower component Spinor $\left\{ \begin{array}{c}
0\\
\psi_{b}\end{array}\right\} $to upper one $\left\{ \begin{array}{c}
\psi_{a}\\
0\end{array}\right\} $. If $\psi$to be an eigen state of $H_{+}(H_{-})$, $\widehat{Q}$$\psi(\widehat{Q}^{\dagger}$$\psi)$is
the eigen state of that of $H_{+}(H_{-})$with equal energy.

Case II- $m\,\neq0$but potential $\Phi=0$. Corresponding Dirac's
equation (\ref{eq:63}) with its solution (\ref{eq:65})is described
as

\begin{eqnarray}
(\gamma_{0}E\,-\gamma_{1}p_{1}-\gamma_{2}p_{2}-\gamma_{3}p_{3}-\, m) & \psi(x)= & 0\label{eq:76}\end{eqnarray}

which may be written as follows in terms of quaternion units i.e.

\begin{eqnarray}
\left\{ \left[\begin{array}{cc}
1 & 0\\
0 & -1\end{array}\right]E-e_{l}\left[\begin{array}{cc}
0 & 1\\
1 & 0\end{array}\right]p_{l}-m\left[\begin{array}{cc}
1 & 0\\
0 & 1\end{array}\right]\right\} \left\{ \begin{array}{c}
\psi_{a}\\
\psi_{b}\end{array}\right\}  & = & 0\label{eq:77}\end{eqnarray}

Accordingly ,we have the following sets of equations

\begin{eqnarray}
\widehat{A}^{\dagger}\psi_{b} & = & (E-m)\psi_{a}\nonumber \\
\widehat{A}\,\psi_{a} & = & (E+m)\psi_{b}\nonumber \\
\widehat{A}^{\dagger}\widehat{A}\,\psi_{a} & = & (E^{2}-m^{2}\,)\psi_{a}\nonumber \\
\widehat{A}\,\widehat{A}^{\dagger}\psi_{b} & = & (E^{2}-m^{2}\,)\psi_{b}\nonumber \\
\widehat{P_{l}}^{2}\psi_{a,b} & = & (E^{2}-m^{2}\,)\psi_{a,b}\label{eq:78}\end{eqnarray}

which are the Schrödinger equation for free particle.SUSY Hamiltonian
is now described as 

\begin{eqnarray}
\widehat{H} & = & \left[\begin{array}{cc}
\widehat{Q}\widehat{Q}^{\dagger} & 0\\
0 & \widehat{Q}^{\dagger}\widehat{Q}\end{array}\right]=\left[\begin{array}{cc}
\widehat{P_{l}}^{2}+m^{2} & 0\\
0 & \widehat{P_{l}}^{2}+m^{2}\end{array}\right]\label{eq:79}\end{eqnarray}

On the similar way we get the following relations while restricting
for two dimensional structure of space and time i.e.

\begin{eqnarray}
\widehat{H} & = & \left[\begin{array}{cc}
\widehat{Q}\widehat{Q}^{\dagger} & 0\\
0 & \widehat{Q}^{\dagger}\widehat{Q}\end{array}\right]=\left[\begin{array}{cc}
-\frac{d^{2}}{dx^{2}}+m^{2} & 0\\
0 & -\frac{d^{2}}{dx^{2}}+m^{2}\end{array}\right]\label{eq:80}\end{eqnarray}

where \begin{eqnarray}
\widehat{Q} & = & \left[\begin{array}{cc}
0 & -e_{2}^{\dagger}\frac{d}{dx}+m\\
0 & 0\end{array}\right]\nonumber \\
\widehat{Q}^{\dagger} & = & \left[\begin{array}{cc}
0 & 0\\
-e_{2}^{\dagger}\frac{d}{dx}+m & 0\end{array}\right]\label{eq:81}\end{eqnarray}

Hence we restore the property of SUSY quantum mechanics and obtain
the commutation and anti commutation relations same that of equation
(\ref{eq:75}) for the free particle Dirac equation with mass as well.

Case III- We now discuss and verify the SUSY quantum mechanics relations
for $m\neq0$ with scalar potential $\Phi=V$.We extend the present
theory in the same manner and express Dirac Hamiltonian in the following
form;

\begin{eqnarray}
\widehat{H_{D}} & = & \left[\begin{array}{cc}
0 & e_{l}p_{l}\\
-e_{l}p_{l} & 0\end{array}\right]+\left[\begin{array}{cc}
0 & -i\,(m+V)\\
i\,(m+V) & 0\end{array}\right]\label{eq:82}\end{eqnarray}

or\begin{eqnarray}
\widehat{H_{D}} & = & \left[\begin{array}{cc}
0 & e_{l}p_{l}-i\,(m+V)\\
-e_{l}p_{l}+i\,(m+V) & 0\end{array}\right]\label{eq:83}\end{eqnarray}

or \begin{eqnarray}
\widehat{H} & = & \left[\begin{array}{cc}
\widehat{Q}\widehat{Q}^{\dagger} & 0\\
0 & \widehat{Q}^{\dagger}\widehat{Q}\end{array}\right]=\left[\begin{array}{cc}
-\frac{d^{2}}{dx^{2}}+(m+V)^{2} & 0\\
0 & -\frac{d^{2}}{dx^{2}}+(m+V)^{2}\end{array}\right]\label{eq:84}\end{eqnarray}

where\begin{eqnarray}
\widehat{Q} & = & \left[\begin{array}{cc}
0 & -e_{2}^{\dagger}\frac{d}{dx}+i\,(m+V)\\
0 & 0\end{array}\right]\nonumber \\
\widehat{Q}^{\dagger} & = & \left[\begin{array}{cc}
0 & 0\\
-e_{2}^{\dagger}\frac{d}{dx}+i\,(m+V) & 0\end{array}\right]\label{eq:85}\end{eqnarray}

Equations\eqref{eq:82} to\eqref{eq:85} satisfy the supersymmetric
quantum mechanical relations given by equation (\ref{eq:75}) and
as such the supersymmetry is verified even for interacting case with
scalar potential.

Case IV- Dirac equation in Electromagnetic Field-

Before writing the quaternion Dirac equation in generalized electromagnetic
fields of dyons let us start with the quaternion gauge transformations.
A $Q-$field (\ref{eq:1}) is described in terms of following SO(4)
local gauge transformations\cite{key-6,key-9};

\begin{eqnarray}
\phi & \rightarrow\phi' & =U\,\phi\,\bar{V\,}\,\,\,\,\,\, U,V\,\varepsilon Q\,\,,\,\,\, U\,\bar{U\,}=V\,\overline{V}=1\label{eq:86}\end{eqnarray}

The covariant derivative for this is then written in terms of two
gauge potentials as 

\begin{eqnarray}
D_{\mu}\phi & = & \partial_{\mu}\phi+A_{\mu}\phi-\phi B_{\mu}\label{eq:87}\end{eqnarray}

where potential transforms as 

\begin{eqnarray*}
A_{\mu}^{'} & = & U\, A_{\mu}\bar{U}\,+U\,\partial_{\mu}\bar{U}\,\end{eqnarray*}
\begin{eqnarray}
B_{\mu}^{'} & = & V\, B_{\mu}\bar{V}\,+V\,\partial_{\mu}\bar{V}\,\label{eq:88}\end{eqnarray}

and

\begin{eqnarray}
\bar{\phi'}\phi' & =\overline{(U\phi\bar{V)}}((U\phi\bar{V)} & =\bar{\phi}\phi=\phi_{0}^{2}+\left|\vec{\phi}\right|^{^{2}}\label{eq:89}\end{eqnarray}

Here we identify the non Abelian gauge fields $A_{\mu}$and $B_{\mu}$as
the gauge potentials respectively for electric and magnetic charges
of dyons described earlier in section-3.Corresponding field momentum
of equation (\ref{eq:87})may also be written as follows

\begin{eqnarray}
P_{\mu}\phi & = & p_{\mu}\phi+A_{\mu}\phi-\phi B_{\mu}\label{eq:90}\end{eqnarray}

where the gauge group $SO(4)=SU(2)_{e}\times SU(2)_{g}$ is constructed
in terms of quaternion units of electric and magnetic gauges .Accordingly,
the covariant derivative thus describes two different gauge field
strengths i.e.

\begin{eqnarray}
\left[D_{\mu},D_{\upsilon}\right]\phi & = & f_{\mu\nu}\phi-\phi h_{\mu\nu}\nonumber \\
f_{\mu\nu} & = & A_{\mu,\nu}-A_{\nu,\mu}+\left[A_{\mu},A_{\nu}\right]\nonumber \\
h_{\mu\nu} & = & B_{\mu,\nu}-B_{\nu,\mu}+\left[B_{\mu},B_{\nu}\right]\label{eq:91}\end{eqnarray}

where $f_{\mu\nu}$ and $h_{\mu\nu}$ are gauge field strengths associated
with electric and magnetic charges of dyons respectively.We may now
write the Dirac equation as

\begin{eqnarray*}
i\,\,\gamma_{\mu\,\,}D_{\mu}\psi(x,t) & = & m\,\psi(x,t)\,\end{eqnarray*}

and accordingly with some restrictions and using the properties of
quaternions we may write the Dirac equation as\begin{eqnarray}
\left[\begin{array}{cc}
m & e_{\mu}(p_{\mu}+A_{\mu}-B_{\mu})\\
-e_{\mu}(p_{\mu}+A_{\mu}-B_{\mu}) & m\end{array}\right]\left[\begin{array}{c}
\psi_{a}\\
\psi_{b}\end{array}\right] & = & E\left[\begin{array}{c}
\psi_{a}\\
\psi_{b}\end{array}\right]\label{eq:92}\end{eqnarray}

where $\varphi_{a}=\varphi_{0}+i\varphi_{1}$and $\varphi_{b}=\varphi_{2}-i\varphi_{3}$and
as such we obtain the following set of equations;\begin{eqnarray*}
E\psi_{a}= & m\psi_{a}+e_{\mu}(p_{\mu}\psi_{b}+A_{\mu}\psi_{b}-\psi_{b}B_{\mu}) & ;\\
E\psi_{b}= & m\psi_{b}+e_{\mu}(p_{\mu}\psi_{a}+A_{\mu}\psi_{a}-\psi_{a}B_{\mu}) & ;\\
e_{\mu}(p_{\mu}\psi_{b}+A_{\mu}\psi_{b}-\psi_{b}B_{\mu}) & = & (E-m)\psi_{a}\\
e_{\mu}(p_{\mu}\psi_{a}+A_{\mu}\psi_{a}-\psi_{a}B_{\mu}) & = & (E-m)\psi_{b}\end{eqnarray*}

\begin{eqnarray}
\widehat{A}^{\dagger}\psi_{b} & = & (E-m)\psi_{a}=e_{\mu}(p_{\mu}\psi_{b}+A_{\mu}\psi_{b}-\psi_{b}B_{\mu})\nonumber \\
\widehat{A}\,\psi_{a} & = & (E+m)\psi_{b}=e_{\mu}(p_{\mu}\psi_{a}+A_{\mu}\psi_{a}-\psi_{a}B_{\mu})\nonumber \\
\widehat{A}^{\dagger}\widehat{A}\,\psi_{a} & = & (E^{2}-m^{2}\,)\psi_{a}\nonumber \\
\widehat{A}\,\widehat{A}^{\dagger}\psi_{b} & = & (E^{2}-m^{2}\,)\psi_{b}\label{eq:93}\end{eqnarray}

where we have restricted ourselves to the case of two dimensional
supersymmetry by imposing the condition $A_{1}^{\dagger}=-A_{1},A_{2}^{\dagger}=-A_{2},A_{3}^{\dagger}=-A_{3},B_{1}^{\dagger}=-B_{1},B_{2}^{\dagger}=-B_{2},B_{3}^{\dagger}=-B_{3}$
to restore the supersymmetry.As such it is possible to supersymmetrize
the Dirac equation for generalized electromagnetic fields of dyons
and we obtain the commutation and anti commutation relations given
by equation(\ref{eq:75}) to verify the supersymmetric quantum mechanics
in this case.

\section{Higher dimensional Supersymmetry}

Quaternion differential operator is defined as 

\begin{eqnarray}
\partial & = & -i\,\partial_{t}+e_{1}\partial_{1}+e_{2}\partial_{2}+e_{3}\partial_{3}\nonumber \\
\overline{\partial} & = & -i\,\partial_{t}-e_{1}\partial_{1}-e_{2}\partial_{2}-e_{3}\partial_{3}\label{eq:94}\end{eqnarray}

which describes

\begin{eqnarray}
\partial\bar{\partial} & = & \partial_{t}^{2}+\partial_{1}^{2}+\partial_{2}^{2}+\partial_{3}^{2}\label{eq:95}\end{eqnarray}

and can be decomposed in to two dimensional from as

\begin{eqnarray}
\partial\partial^{\dagger}=-\nabla^{2} & = & (\frac{\partial}{\partial x}+e_{1}\frac{\partial}{\partial y})(-\frac{\partial}{\partial x}+e_{1}\frac{\partial}{\partial y})\label{eq:96}\end{eqnarray}

where ($\dagger$) changes only complex quantities with one quaternion
units which plays the role of complex quantity (in C(1,i) case ) and
thus is equivalent to 

\begin{eqnarray}
-\nabla^{2} & = & (\frac{\partial}{\partial x}+i\frac{\partial}{\partial y})(-\frac{\partial}{\partial x}+i\frac{\partial}{\partial y})\label{eq:97}\end{eqnarray}

Now defining $q=(\frac{\partial}{\partial x}+i\frac{\partial}{\partial y})$
and $q^{\dagger}=(-\frac{\partial}{\partial x}+i\frac{\partial}{\partial y})$we
have the negative of Laplacian $-\nabla^{2}=q\, q^{\dagger}=q^{\dagger}q$andthus
describes two-dimensional free particle Supersymmetric quantum mechanics.
Following Das et al\cite{key-21} we may now construct a two-dimensional
supersymmetric theory in the following manner,

\begin{eqnarray}
q & = & \overrightarrow{a}\cdot\overrightarrow{(\nabla}+\overrightarrow{W}\,)\label{eq:98}\end{eqnarray}

where $\overrightarrow{W}\:$ is described as super potential, $\overrightarrow{a}=a_{x}+ia_{y}$
and $\overrightarrow{\nabla}$represents the two dimensional gradient.We
may now write the super partner hamiltonians described in the previous
sections as\cite{key-21};

\begin{eqnarray}
H_{1} & =q^{\dagger}q & =\sum_{j=1}^{2}(-\nabla_{j}+W_{j})(\nabla_{j}+W_{j})-i\,\epsilon^{jk}\left\{ \nabla_{j},\nabla_{k}\right\} \nonumber \\
H_{2} & =qq^{\dagger} & =\sum_{j=1}^{2}(\nabla_{j}+W_{j})(-\nabla_{j}+W_{j})-i\,\epsilon^{jk}\left\{ \nabla_{j},\nabla_{k}\right\} \label{eq:99}\end{eqnarray}

as described earlier here also the curly brackets represent anti commutators
and $\epsilon^{jk}$ is anti symmetric with $\epsilon^{12}=1$ and
$\epsilon^{21}=-1$.It is to be noted that the vector super potential
naturally generates a gauge field interaction structure resulting
from supersymmetry algebra.Thus it is possible to say that we can
take quaternion supersymmetry as N = 4 real supersymmetry because
like complex quanties generat two dimensional real representation,accordingly
quaternions generate four dimensional real representations. thus we
need to define q in such a manner that $-\nabla^{2}=q\, q^{\dagger}=q^{\dagger}q$
and 4-dimensional supersymmetry algebra can be built accordingly in
terms of three non commutating quantities but associative quantities
like three quaternion units. ej. Let us assume that $q$ in free space
can be written as a linear super position in terms of pure quaternion
units consisting non-Abelian gauge structure and are obtained from
$q$ with $q=\frac{1}{2}(q-\overline{q})$, i.e.

\begin{eqnarray*}
q & = & {\scriptstyle {\displaystyle \sum_{j=1}^{3}e_{j}\nabla_{j}}}\end{eqnarray*}

and we may write equation (\ref{eq:99}) as 

\begin{eqnarray}
H_{1} & = & q^{\dagger}q=\sum_{j=1}^{3}e_{j}e_{j}^{\dagger}(-\nabla_{j}^{2})-\sum_{j<k}^{3}(e_{j}e_{k}^{\dagger}+e_{k}e_{j}^{\dagger})\nabla_{j}\nabla_{k}\nonumber \\
H_{2} & = & qq^{\dagger}=\sum_{j=1}^{3}e_{j}^{\dagger}e_{j}(-\nabla_{j}^{2})-\sum_{j<k}^{3}(e_{j}^{\dagger}e_{k}+e_{k}^{\dagger}e_{j})\nabla_{j}\nabla_{k}\label{eq:100}\end{eqnarray}

where we may write $q=\overrightarrow{a}\cdot\overrightarrow{\nabla}$with
${\scriptstyle {\displaystyle \overrightarrow{a}=\sum_{j=1}^{3}e_{j}a_{j}}}$.The
vector super potential depends on the position as

\begin{eqnarray*}
q=\overrightarrow{a}\cdot\overrightarrow{(\nabla} & + & \overrightarrow{W)}=\sum_{j=1}^{3}e_{j}(\nabla_{j}+W_{j})\end{eqnarray*}

and accordingly we may obtain the super partner Hamiltonians $qq^{\dagger}$and
$q^{\dagger}q$ interms of interacting super potential \cite{key-21}
with the assumption that the gauge interaction structure naturally
arises from the requirement of super symmetry in ters of quaternions
and gauge theory of dyons be dealt in this manner.The generalization
of this theory to octonion is not possible because of the non associative
nature of octonions.Secondly octonions can not be written directly
in terms of eight dimensional matrix representation of real numbers
like quaternions are written in terms four dimensional representation
of real numbers. There is the difference between quaternion and octonions
that the quaternion satisfies all the property of matrices while the
octonions are not and the alternativity property of split octonions
is still not competant to resolve these inconsistencies associated
with octonions. So, it is hard to write the super symmetric extension
and we will have to write it another way to visualize the supersymmetry
with octonions and octonion gauge theory of dyons.thus we may conclude
that many of the properties of quaternion quantum mechanics lead to
the properties of supersymmetric quantum mechanics because in both
the cases the energy is positive semi definite and a gauge interaction
arises automatically when one defines the quaternion units in terms
of Pauli spin matrices and supersymmetric charges are defined accordingly.
If we describe the well known N = 4 supersymmetry, the 4-dimensional
real anti symmetric matrices $\alpha_{j}$and $\beta_{J}$ for all
(j = 1,2,3) associated with it \cite{key-26}satisfy the algebra given
by 

\begin{eqnarray}
\left\{ \alpha^{i},\,\alpha^{j}\right\}  & = & \left\{ \beta^{i},\,\beta^{j}\right\} =-2\,\delta^{ij}\,\,\,\,\,\,\,\,\,\,\,\,\,\,\,\,\,\left[\alpha^{i}\,,\,\beta^{j}\right]=0\nonumber \\
\left[\alpha^{i},\,\alpha^{j}\right] & = & -2\,\varepsilon^{ijk}\alpha^{k}\,\,\,\,\,\,\,\,\,\,\,\,\,\,\,\,\,\,\,\,\,\,\,\,\,\,\,\,\left[\beta^{i},\,\beta^{j}\right]=-2\,\varepsilon^{ijk}\beta^{k}\label{eq:101}\end{eqnarray}

which is the algebra of quaternions. The matrices $\alpha_{j}$ and
$\beta_{J}$are thus the quaternion analogue for give $4\times4$
real matrix representation of three quaternion units since the properties
of $\alpha_{j}$ and $\beta_{J}$ are same as those for three non
abelian quaternion units. Thus we conclude that N=4 real Super symmetry
can be visualized as N=1 quaternion and N=2 complex Super symmetry
and the theories of monopoles and dyons are thus be understood better
in terms of hyper complex number system.

\end{document}